# Asymmetric-detection time-stretch optical microscopy (ATOM) for ultrafast high-contrast cellular imaging in flow

Terence T. W. Wong,[1,2,*] Andy K. S. Lau,[1,*] Kenneth K. Y. Ho,[3,4] Matthew Y. H. Tang,[3] Joseph D. F. Robles,[5] Xiaoming Wei,[1] Antony C. S. Chan,[1] Anson H. L. Tang,[1] Edmund Y. Lam,[1] Kenneth K. Y. Wong,[1] Godfrey C. F. Chan,[5] Ho Cheung Shum,[3] Kevin K. Tsia[1]

[1]Department of Electrical and Electronic Engineering, The University of Hong Kong, Pokfulam Road, Hong Kong, [2]Current address: Department of Biomedical Engineering, Washington University in St. Louis, St. Louis, MO 63130, USA, [3]Department of Mechanical Engineering, The University of Hong Kong, Pokfulam Road, Hong Kong, [4]Current address: Department of Mechanical Engineering, University of Michigan, Ann Arbor, MI 48109, USA, [5]Department of Pediatrics and Adolescent Medicine, Li Ka Shing Faculty of Medicine, The University of Hong Kong Pokfulam Road, Hong Kong, *These authors contributed equally to this work. Correspondence and requests for materials should be addressed to Kevin. K. Tsia (email: tsia@hku.hk).

**Abstract:** Accelerating imaging speed in optical microscopy is often realized at the expense of image contrast, image resolution, and detection sensitivity – a common predicament for advancing high-speed and high-throughput cellular imaging. We here demonstrate a new imaging approach, called asymmetric-detection time-stretch optical microscopy (ATOM), which can deliver ultrafast label-free high-contrast flow imaging with well delineated cellular morphological resolution and in-line optical image amplification to overcome the compromised imaging sensitivity at high speed. We show that ATOM can separately reveal the enhanced phase-gradient and absorption contrast in microfluidic live-cell imaging at a flow speed as high as ~10 m/s, corresponding to an imaging throughput of ~100,000 cells/sec. ATOM could thus be the enabling platform to meet the pressing need for intercalating optical microscopy in cellular assay, e.g. imaging flow cytometry – permitting high-throughput access to the morphological information of the individual cells simultaneously with a multitude of parameters obtained in the standard assay.

High-throughput measurement or screening is routinely utilized in clinical diagnostics and basic research in life sciences, such as pathology, drug discovery, aberrant cells screening in stem cell research, rare cancer cell detection, and emulsion droplets/particle synthesis[1-3]. Very often, these applications demand for enumeration and characterization of large population of specimens (e.g. >100,000 particles) with a high-degree of statistical accuracy. Yet, current approaches have largely been restricted by a trade-off between throughput and accuracy. This is particularly exemplified in the context of cellular assay (or cell-based assay), which is a valuable tool for studying cellular characteristics and dynamics by measuring a multitude of parameters (usually based on fluorescent and scattered light intensity signals) from each cell in a sizable population. In most scenarios, improved measurement accuracy comes with the ability of gaining access to the morphological information of the cells, i.e. to capture images of the cells – facilitating better cellular identification/discrimination and thus yielding high-confidence statistical data. However, the acquisition of more spatial information would inevitably lower the measurement throughput, or vice versa. For example, some cell-based assays allow measurement of multiple parameters per cell with a very high throughput, but without spatial information of each cell. Some others, on the other hand, can capture high-resolution cellular images but with low throughput. This is because of the intrinsic trade-off between imaging speed and sensitivity in the image sensors of the optical microscope systems[4,5]. This explains why the current state-of-the-art imaging flow cytometers can only reach an imaging throughput ~1000 cells/sec, compared with the throughput of ~100,000 cells/sec of the classical non-imaging flow cytometers[3,6,7].

Realizing image-based cellular assays, which provide access to detailed morphological information of the individual cells and enable delivery of multiparametric cytometry without sacrificing the throughput, is hence of significant value. Such high-throughput combinatorial and complementary measurements would particularly benefit accurate rare cells detection and single-cell analysis within a large single or heterogeneous population of cells[8-10].

In this regard, a new optical imaging modality called time-stretch microscopy, which bypasses the intrinsic limitation of the image sensor, has recently been developed for ultrafast imaging. It has been found particularly pertinent to microparticle imaging in microfluidic flow, with an imaging throughput comparable to conventional non-imaging flow cytometry[11]. It is achieved by ultrahigh-speed retrieval of image information (at MHz line-scan rate or frame rate) encoded in the spectrum of a broadband and ultrashort optical pulse (femtoseconds to picoseconds) by converting it into a serial temporal data in real time. This technique has proven to be able to perform high-throughput image-based cancer cell screening, which is an ideal complementary tool to typical multiparametric flow cytometry[12]. However, time-stretch microscopy has so far mostly been operated in bright-field (BF) imaging mode in longer wavelength range [11,12], and is thus not capable of revealing high-contrast and detailed morphology of the transparent cells – hindering accurate cell recognition and screening. As a result, effective use of time-stretch imaging to-date only has been limited to microparticle or cell screening in high-speed flow with trivial size and shape differences, especially when the targeted cells are labeled with contrast agents[12]. Similar to the classical optical imaging modalities, the time-stretch image quality, which is typically characterized by image resolution and image contrast[13], is still compromised by the higher imaging speed. With the aim of enlarging the scope of time-stretch imaging applications, we here introduce a new imaging approach called *asymmetric-detection time-stretch optical microscopy* (ATOM), for obtaining label-free, high-contrast image of the transparent cells at ultrahigh speed, and with sub-cellular resolution.

The central idea is to generate the enhanced phase-gradient contrast in the time-stretch image based on a simple asymmetric detection scheme (Fig. 1). The phase-gradient contrast results in three-dimensional (3D) appearance in the image, resembling the contrast-enhancement effect in Schlieren imaging[14,15]. Moreover, by time-multiplexing two ATOM images with opposite phase-gradient contrasts, we further obtain two different contrasts from the same specimen: one with differential (enhanced) phase-gradient contrast and another with absorption contrast, simultaneously. This method decouples the phase-gradient information from absorption, resulting in further enhancement of the image contrast[16-18]. The time-multiplexing scheme does not significantly compromise the final imaging speed, because of the intrinsically high-speed operation of time-stretch (> tens of MHz). Together with operating in the 1 μm wavelength regime, a more favorable spectral window for biophotonic applications as opposed to the telecommunication wavelength band (~1.5μm) used by most of the previously reported time-stretch imaging systems, the ATOM system presented here is able to achieve higher diffraction-limited resolution and high-contrast cellular time-stretch imaging. Similar to the original time-stretch imaging, the detection sensitivity in ATOM is not compromised by the high-speed operation because of the in-line *optical image amplification* – a feature rarely implemented in typical optical microscopy. We demonstrate the unique capability of ATOM in visualizing the detailed cellular morphology (e.g. normal human blood cells from fresh blood and human leukemic monocytes) without contrast agent in ultrafast microfluidic flow (up to ~10 m/s), which is yet to be demonstrated in the existing time-stretch imaging modality. The achieved flow speed here is equivalent to an imaging throughput of ~ 100,000 cells/sec. ATOM thus represents a significant advancement in bringing the essential imaging metrics – high resolution and high contrast – to high-speed time-stretch imaging, making it a genuinely appealing platform for realizing high-throughput image-based cellular assay.

## Results

**General working principles.** In its original configuration, time-stretch imaging is accomplished by a two-step signal mapping process: the spatial information of the specimen is first mapped to the spectrum of a broadband laser pulse by using the diffractive optical element (diffraction grating in our case, as shown in Fig. 1). The encoded spectrum of the pulse is then mapped (stretched) into a serial temporal data format in real time via group velocity dispersion (GVD) in a dispersive optical fiber. The time-stretch pulse, now encoded with spatial information of the image, is captured by a high-speed single-pixel photo-detector instead of the image sensors (Fig. 2(a))[11]. In contrast, ATOM further creates phase-gradient contrast of the time-stretch images by asymmetrically detecting the spectrally-encoded pulses prior to the time-stretch process. It is done by off-axis coupling of the encoded pulsed beam into the dispersive fiber core, which essentially acts as the confocal pinhole of the ATOM system (Fig. 1). By introducing an oblique angle $\theta$ between the fiber axis and the laser beam propagation axis, the cone of the coupling light becomes asymmetric. In effect, it is equivalent to partially blocking the beam detection path – a common approach taken in Schlieren imaging. Such detection scheme results in images with phase-gradient contrast, which has similar characteristic 3D appearance in differential interference contrast (DIC) microscopy[16,18]. Compared to classical phase-contrast and DIC microscopy, the contrast enhancement mechanism in ATOM does not rely on interference. Instead, it reveals the phase-gradient contrast by detecting the wavefront tilt through a simple asymmetric detection. It also requires no additional polarizing optics to generate the differential phase contrast, as in the case of DIC microscopy. Therefore, ATOM, operating at an imaging speed of orders-of-magnitude faster than that of ordinary DIC microscopy, represents a simple but robust technique of realizing ultrafast high-contrast optical microscopy, and particularly benefits cellular flow imaging applications.

In ATOM, all the encoded wavelengths are recombined by the diffraction grating and are collected by the dispersive fiber within the same asymmetric coupling light cone. Hence, the same phase-gradient contrast for all wavelengths is preserved (Fig. 1). Note that we choose the asymmetric detection scheme over asymmetric illumination, which can also yield phase-gradient contrast[16,18], because it requires no modification in the original imaging path of the time-stretch microscope, and thus provides more flexibility to optimize the image contrast by simply adjusting the off-axis fiber coupling angle $\theta$. The optical loss due to the off-axis coupling misalignment is not a critical issue in ATOM because it can readily be compensated by the optical gain provided by fiber amplification in-line with the time-stretch process[11].

The general schematic of an ATOM system is shown in Fig. 2(a). It consists of standard spectral shower illumination for time-stretch imaging[11,12,19-22], which is operated in a double-pass transmission mode (see Method). The double-passed spectral shower, which encodes the information of the sample, is then restored back to original pulsed beam by the grating and is collected by a dispersive fiber using a fiber collimator lens (see Method). This pulsed beam is coupled into the fiber with an adjustable off-axis angle $\theta$ in order to fine-tune the asymmetric detection condition, and hence the phase-gradient contrast of the final image. The spectrally-encoded pulse is then mapped to the time within the dispersive fiber with GVD of 0.35 ns/nm. The time-stretch signal is also amplified by an in-line fiber-based semiconductor optical amplifier (SOA) which achieves an on-off gain as high as ~100, in order to compensate the off-axis fiber coupling loss as well as the dispersive loss in the fiber. Without such optical (image) gain, the time-stretch signal is too weak to be detectable (see Supplementary Fig. S1). Finally, the signal is detected by a photo-detector (bandwidth: 10 GHz) and a real-time oscilloscope (sampling rate: 40 GS/s). With the high GVD as well as the high-bandwidth oscilloscope, our current system achieves the diffraction-limited image resolution at ~1.2 μm[21,23].

We stress that a wide range of choices of optical amplification can be used in-line with ATOM operated in the 1-μm wavelength regime. They can be either discrete amplifiers (namely SOA, ytterbium-doped fiber amplifier, or optical parametric amplifier) or distributed amplifiers, such as distributed fiber Raman amplifier[11,20,24]. Optimal choice depends on the required gain bandwidth, total gain, as well as noise figure of the amplifiers[24,25]. Note that the GVD achieved is high enough to ensure the final image resolution is diffraction-limited, and is not limited by the GVD or the detector's bandwidth[21,23].

Furthermore, one can access different phase-gradient contrasts simultaneously within each line scan (i.e. each laser repetition period) by time-multiplexing more than one time-stretch signal. An interesting configuration is to generate two time-multiplexed, i.e. time-delayed, replicas which give opposite phase-gradient contrasts, from the same line scan (see Method and Supplementary Fig. S2). This is simply done by first splitting the spectrally-encoded pulsed beam into two paths – one is time-delayed with respect to the other. The two beams are then launched into the same dispersive fiber with opposite coupling orientations with respect to the fiber axis (Fig. 2(a)). Consequently, we can simultaneously obtain two separate time-stretch images of the same specimen, with different contrasts in a single-run of ATOM measurement: one with *differential* phase-gradient contrast (subtraction of the two signals) and another with absorption contrast (addition of the two signals) (Fig. 2(b)). This method thus allows the absorption information of the specimen decoupled from the phase-gradient information[16,17]. It should be emphasized that such time-multiplexing scheme results in no compromise on the imaging speed as long as the total duration of two replicas does not exceed one line-scan period (i.e. total duty cycle should be kept < 100%). Thus, ultrafast operation can be maintained, i.e. at the line-scan rate of >MHz), as shown in Supplementary Fig. S2.

**Basic performance of ATOM.** We first performed ATOM of the unstained immortalized normal hepatocyte cells (MIHA) fixed on a glass slide, with two different coupling orientations. In this case, two-dimensional (2D) ATOM images are acquired by scanning the sample stage orthogonal to the spectral shower direction at a single-shot line-scan rate of 1 MHz (see details in Method). By tuning the fiber coupling angle above and below the fiber axis, we can capture the ATOM images with two opposite phase-gradient contrasts, i.e. the shadow cast can be switched to the opposite side of the MIHA cell (Fig. 3(a) and (b), see also the corresponding line profile comparison between Fig. 3(c) and (d)). Computing the sum and difference of these two opposite-contrast ATOM images allows us to simultaneously obtain both absorption and differential (enhanced) phase-gradient-contrast images of the MIHA cell, respectively (Fig. 3(e) and (f)). The enhanced phase-gradient contrast revealed by ATOM closely resembles that obtained by the white-light DIC microscopy (Fig. 3(g)) but with ultrafast imaging speed, i.e. each line scan is as short as ~4 ns.

We also performed ATOM of the MIHA cells in ultrahigh speed flow in a polydimethylsiloxane (PDMS) microfluidic channel to demonstrate the contrast enhancement of ATOM compared to ordinary time-stretch microscopy (BF mode, i.e. on-axis fiber coupling). The channel is designed in a way that the balance between the inertial lift force and the viscous drag force is achieved for manipulating the positions of the individual cells and focusing them in high-speed flow[26,27]. This microfluidic technique is essential for ensuring robust imaging by ATOM at the record high microfluidic flow speed, as high as ~10 m/s, which is limited only by the pressure that the channel can withstand (see Method, Supplementary Figs. S3 and S4 for detailed design and fabrication steps). Note that this flow speed corresponds to an imaging throughput up to ~100,000 cells/sec, which is orders-of-magnitude higher than any existing imaging flow cytometers[3]. Here, 2D images are acquired by continuous 1D line-scan at a rate of 26 MHz (governed by the repetition rate of the laser), which is naturally provided by the unidirectional cell flow, without any laser beam or sample stage scanning. The enhanced contrast in ATOM images compared to the BF time-stretch images is clearly evident in Fig. 3(h) and (i) (see more examples in Supplementary Fig. S5) – enabling visualization of detailed cellular morphology under such an ultrahigh flow speed. Hence, it demonstrates the obvious advantage of ATOM over BF time-stretch – accessing phase gradient of the specimen to boost the time-stretch imaging contrast without sacrificing the speed (maintaining at tens of MHz). It should be noted that the ability to perform high (phase-gradient) contrast microscopy in an exceptionally high flow speed (up to ~10 m/s) is unique to ATOM and it could not be readily accomplished by any state-of-the-art high-speed camera (see the image captured by a high-speed CMOS camera, as shown in Supplementary Fig. S5(b)).

**High-contrast cellular microscopy in ultrafast flow by ATOM.** We further demonstrate ultrafast and high-contrast ATOM by imaging stain-free human hepatocellular carcinoma cells (BEL-7402) and human cervical cancer cells (HeLa), flowing at an ultra-high speed of 7-8 m/s in the same microfluidic channel. The representative ATOM images are shown in Fig. 4 (More images are shown in Supplementary Figs. S6 and S7). The images were captured by averaging every 4 single-shot line-scans in order to further improve the signal-to-noise ratio, resulting in an effective line-scan rate of ~6.5 MHz. The ATOM images have negligible image blur even at such a high flow speed, thanks to its ultra-short exposure time of ~20ps, which is governed by the time-bandwidth product of the each spectrally-resolvable subpulse of the spectral shower[11, 21]. Such exposure time for *real-time* imaging is essentially unattainable by any existing image sensors. More importantly, the ATOM reveals enhanced time-stretch image contrast, particularly apparent in the differential phase-gradient ATOM images (Figs. 4(b) and (e)), in which the absorption information has been separated (see Figs. 4(c) and (f)).

Moreover, high-contrast (differential phase-gradient-contrast) ATOM allows unambiguous imaging of the individual cells in an aggregate, and thus enables clear identification of such aggregate event in a high-speed flow (Figs. 4(b) and (e)). We note that conventional non-imaging flow cytometers are incapable of identifying such aggregated-cell/cluster events, which are often falsely regarded as a large single entity. Consequently, these aggregated cells would easily be missed by standard forward-scatter or side-scatter gating analysis – resulting in erroneous statistical results[2,6]. Showing the ability to visualize single cells with high contrast and in high speed, the results presented here thereby concisely demonstrates the unique strength of ATOM for realizing imaging flow cytometry with high throughput (~100,000 cells/sec) as well as high accuracy.

Finally, we captured the ATOM images of the acute monocytic leukemia cells (THP-1) as well as normal human blood cells, from fresh blood obtained from a healthy donor, flowing in the microfluidic channel at a speed as high as 10 m/s. It is worth noting that the differential phase gradient contrast in the ATOM images of the THP-1 cells in general enables us to visualize the nuclei (Figs. 5(b) and (d), and more images are shown in Supplementary Fig. S8) – imaging such sub-cellular structure without label is yet to be demonstrated in ordinary BF time-stretch imaging. In addition, ATOM is also able to identify the biconcave disk shape of the red blood cells (RBCs) (Fig. 5(g)) and can differentiate them from the swelled RBCs in the flow, which are in either spherical or elliptical shapes (Fig. 5(h)). More images are shown in Supplementary Fig. S9. These different morphologies are confirmed by comparing ATOM images with the ordinary DIC images of the same blood sample (see Supplementary Fig. S9).

## Discussion

We have demonstrated a new ultrafast optical imaging technique, in the context of imaging in flow, called ATOM, which delivers high-contrast single-cell imaging in an ultrahigh-speed microfluidic flow up to 10 m/s – orders-of-magnitude faster than the typical optofluidic cellular imaging techniques (~ mm/s)[28]. Although time-stretch optical microscopy has been developed and has targeted high-throughput cell screening applications, its BF imaging mode, which results in low image contrast, hinders accurate image-based cellular identification. Inevitably, low false-positive rate in the statistical measurement cannot be guaranteed, albeit its unusually fast imaging speed. Without resorting to exogenous contrast agent, ATOM exploits label-free/stain-free phase-gradient contrast of the cells by a simple and robust asymmetric detection scheme. Thanks to its ultrafast line-scan rate (> tens of MHz), ATOM can further access both the differential (enhanced) gradient-phase contrast and the absorption contrast simultaneously through time-multiplexing two spectrally-encoded pulses – without compromising the final imaging speed.

We also stress that the ATOM system reported here is operated in the shorter near infrared window, ~1 µm, in contrast to most of the prior work on time-stretch imaging which operates in the telecommunication band, primarily because of the wide availability of the low-loss dispersive fibers – the key element for the time-stretch process. By exploiting proper dispersive elements (e.g. specialty dispersive fiber and few-mode fibers[21,22], we here demonstrate that high GVD (~0.35 ns/nm), and thus high-resolution time-stretch imaging, can also be feasible in the 1µm window. It is not only favorable for imaging with higher diffraction-limited resolution, but also opens up a wider scope of biophotonic applications, especially for in-vivo imaging.

The unique feature of ATOM, i.e. the enhanced phase-gradient image contrast, together with the higher diffraction-limited resolution, permits accurate image-based cellular assay with genuinely high throughput. This is particularly exemplified by its capability of visualizing the individual mammalian cells in an aggregate under high-speed flow (Fig. 4), which is otherwise easily missed in standard non-imaging flow cytometry as they do not have access to spatial information of the cells – diminishing the measurement accuracy and reliability. We have also further demonstrated the feasibility of imaging stain-free normal human blood cells as well as leukemic cells by ATOM, where the characteristic cellular features of these individual blood cells are clearly discerned.

Enabling high-contrast cellular imaging in conventional flow cytometry allows accurate cell population identification and classification without substantially relying on subjective manual partitioning (or gating) of cell events. Such gating method is universal in flow cytometry data analysis, but commonly results in misinterpretation of the collected statistical data[29,30]. The results of ATOM presented here are of great significance for advancing time-stretch imaging for high-throughput imaging flow cytometry with high statistical precision. It is particularly envisaged for rare cell screening in early metastasis detection, or post-chemotherapy detection of the residual cancer cells, a concept called minimal residual disease (MRD) detection[31]. The image processing and reconstruction of ATOM is currently done off-line, and the total data size is limited by the memory capacity of the real-time oscilloscope. Real-time ATOM and automated image analysis could be made possible by integrating the system with parallel digital signal processing based on field-programmable gate array (FPGA)[12] or graphic processing unit (GPU). As a result, it would be a powerful tool for high-throughput image-based cellular assay, particularly complementary to the multiparametric analysis of the existing non-imaging flow cytometry.

## Methods

**Illumination and imaging optics of ATOM.** The pulsed beam of a home-built ytterbium-doped mode-locked laser (repetition rate = 26 MHz; center wavelength = 1064 nm) with a 3-dB bandwidth of ~10 nm and a pulse width of 4 ps is spatially dispersed in 1D by a transmission holographic grating (1200 lines/mm) to generate a spectral shower which is then focused by an objective lens (numerical aperture (NA) = 0.66) for illumination. Another identical objective lens (NA = 0.66) and a mirror are added behind the sample in order to operate ATOM in a double-pass transmission mode (Fig. 2(a)). The double-passed spectral shower is then collected by a dispersive fiber using a fiber collimator lens (NA = 0.25).

**Time multiplexing the time-stretch temporal signals.** The double-passed spectrally-encoded light is split into two different paths by a beam splitter (power splitting ratio = 45:55). One of the beams (beam A in Fig. 2(a)) is incident to the fiber collimator lens with an angle range of +4°, whereas the other beam is time-delayed by ~3.6 ns with respect to beam A and is incident to the same fiber collimator lens with an angle range of –4° (beam B in Fig. 2(a)) The tilting angles of the two beams are independently controlled by the two steering mirrors, as shown in Fig. 2(a). The imaging region of interest is mainly near the center of the microfluidic channel, within a size of ~30 μm (i.e. the cell size in most of our experiments), which corresponds to a wavelength bandwidth of ~5 nm and thus a temporal width of ~1.75 ns (with GVD = 0.35 ns/nm). Therefore, a time delay of 3.6 ns guarantees no temporal overlap between the two pulses. The pulses are then time-stretched within a dispersive fiber module, consisting of a 5-km single-mode fiber (SMF) in 1μm and a standard telecommunication SMF (SMF28), which acts as a few-mode fiber[22]. The total GVD achieved is ~0.35 ns/nm. The time-stretch pulses are also amplified by a fiber-based SOA which achieves an on-off gain as high as ~100. Finally, the signal is detected by a photo-detector (bandwidth: 10 GHz) and a real-time oscilloscope (sampling rate: 40 GS/s).

**Image processing of ATOM.** The individual time-stretch waveforms are first subtracted and normalized by the background pulse (which has the spectral shape of the laser source). The final 2D ATOM image is obtained by digitally stacking all the pulses, i.e. the line scans, along the flow direction. The differential phase-gradient and absorption contrasts are further obtained by calculating the difference and sum between the two neighboring time-stretch waveforms, (having opposite phase-gradient contrasts), respectively. The digital signal processing and image reconstruction are done off-line by a custom program in MATLAB.

**Imaging the fixed MIHA cells with ATOM.** For the ATOM images shown in Fig. 3, the MIHA cells are fixed on a glass slide, which is scanned perpendicular to the spectral shower direction during imaging by ATOM. The MIHA sample is scanned for 70 lines with 1 μm step size by a mechanical scanning stage. 2D images are obtained by digitally stacking the 1D line-scans. Averaging of 25 sequential line-scans is taken for realizing an effective single-shot line-scan rate of 1MHz.

**Microfluidic channel design and fabrication.** We designed and fabricated a polydimethylsiloxane (PDMS) microfluidic channel platform in which the balance between the inertial lift force and the viscous drag force is achieved for manipulating the positions of the individual cells and focusing them in ultrafast flow inside the channel. This microfluidic technique is essential for ensuring robust imaging by ATOM at the record high microfluidic flow speed (as high as ~10 m/s). Detailed fabrication steps are depicted in Supplementary Fig. S3. The microfluidic platform consists of two parts: a focusing section followed by an imaging section (see Supplementary Fig. S4(a)). The focusing section consists of multiple pairs of connected curved channels with radii of curvature 400 µm and 1000 µm, respectively. There are 16 turns in total. (see Supplementary Fig. S4(a)). The width (150 µm) and height (50 µm) of the channel were chosen such that the channel is suitable for focusing cells with a size ranging from ~5 - 20 µm. In the imaging section in which the spectral shower is illuminated onto the channel, the channel width is narrowed to 45 µm to further boost up the flow speed. Note that laminar flow condition is still satisfied at such an ultrafast flow. The Reynolds number of our current microfluidic channel design is at 600, which is far below the limit of 2000, beyond which the turbulence flow occurs[32]. The thicknesses of the top and bottom channel walls have been minimized to accommodate the high NA objective lens, which typically has a working distance of less than 1 mm (see Supplementary Fig. S4(b)).

**Preparation of mammalian cell lines – MIHA, BEL-7402 and HeLa cells.** The BEL-7402 cells, HeLa cells, and MIHA cells were cultured on 100mm cell culture dish (Corning) in DMEM-HG supplemented with 10% fetal bovine serum (FBS), 100 U/ml penicillin and 100 μg/ml streptomycin. Sodium pyruvate was added to the culture media when culturing MIHA cells. Cells were grown in a humidified incubator at 37 °C and 5% CO2. After reaching confluence, cells in cell culture dish were trypsinized to suspend in aqueous environment. Trypsin was removed by centrifuging the cell sample at 250g for 5 mins. Glutaraldehyde was added to re-suspend the cells and fix the cells. Glutaraldehyde was removed by centrifuging the cell sample by 250 g for 5 mins. Phosphate buffered saline (PBS) was added to re-suspend the cells which are then loaded to the microfluidic channel for ATOM experiments.

**Preparation of cell lines (THP-1), and human whole blood samples.** THP-1, a human monocytic cell line obtained from patients with acute monocytic leukemia, was cultured in Roswell Park Memorial Institute (RPMI) medium 1640 (Gibco) supplemented with 10% fetal bovine serum (Hyclone, Thermo scientific), penicillin streptomycin (Gibco), 2mM Glutamax$^{TM}$ (Gibco) at 37 °C, 95% humidity and 5% $CO_2$ until confluence was achieved. Optimal conditions were maintained until the cells were utilized for the experiment. Fresh blood was collected from the right median cubital artery of a healthy donor amounting to 3 mililitres and kept at 2-8 °C in an ethylene diamine tetraacetic acid (EDTA) anticoagulated evacuated tube (Greiner Bio-one). Blood was drawn at least eight hours prior to the experiment.

**Acknowledgements**

We thank Hilary K. Y. Mak for preparing the MIHA, BEL-7402 and HeLa cell lines for us. This work was partially supported by grant from the Research Grants Council of the Hong Kong Special Administrative Region, China (Project No. HKU 7172/12E, HKU 717510E, HKU 717911E, HKU 720112E) and University Development Fund of HKU.

**Author contributions**

The ATOM idea was conceived by T.T.W.W. The imaging scheme was designed and developed by T.T.W.W., A.K.S.L. and K.K.T. The experiments were performed by T.T.W.W. and A.K.S.L. The microfluidic channel was designed by T.T.W.W., K.K.Y.H. and M.Y.H.T. The microfluidic channel was fabricated by K.K.Y.H. and M.Y.H.T. The laser source was developed by X.M.W. The data analyses

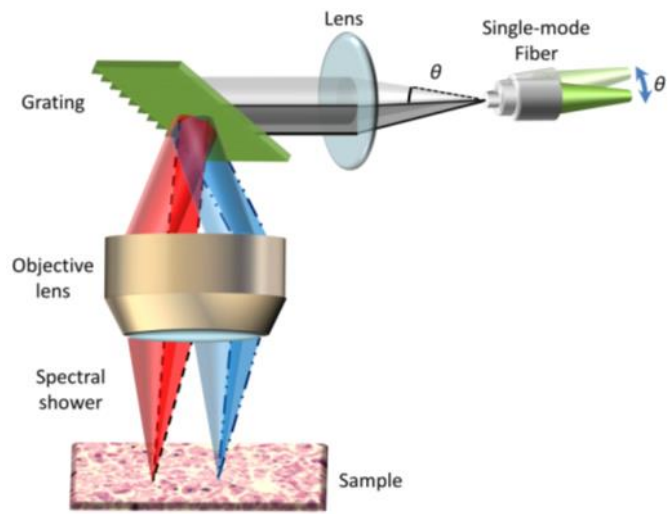

**Figure 1 | Key approach of enabling phase-gradient contrast in ATOM**. In a typical configuration of time-stretch optical microscopy, the spatial information of the specimen is first mapped to the spectrum of a broadband laser pulsed beam by using a diffraction grating through bright-field illumination. The spectrally-encoded pulsed beam is then coupled on-axis into the fiber core of the dispersive fiber, which acts as the confocal pinhole of the imaging system. In contrast, the spectrally-encoded pulsed beam is coupled off-axis into the fiber core in ATOM. By introducing an oblique angle $\theta$ between the fiber axis and the beam propagation axis, the cone of the coupling light becomes asymmetric (darker gray area of the coupling beam shown in the figure). In effect, it is equivalent to partially blocking the beam detection path, and thus to asymmetrically capturing the light from the sample – giving rise the phase-gradient contrast. As all the encoded wavelengths (e.g. blue and red components depicted in the figure) are recombined by the same diffraction grating and are collected by the fiber within the same asymmetric coupling light cone. Hence, same degree of phase-gradient contrast for all wavelengths is preserved (see the areas enclosed by the dashed lines of both the red and blue components).

(a)

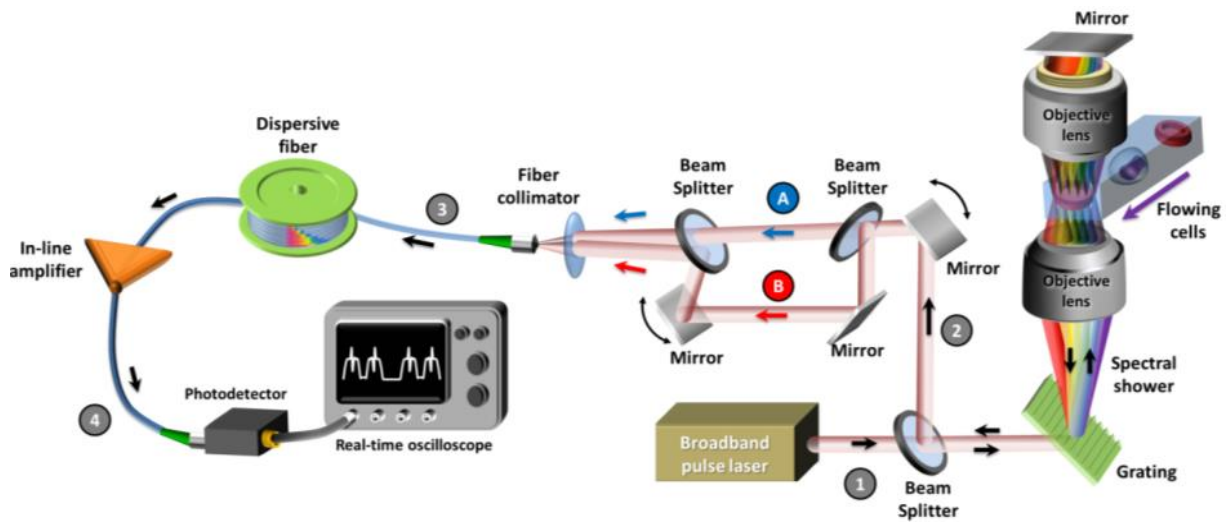

(b)

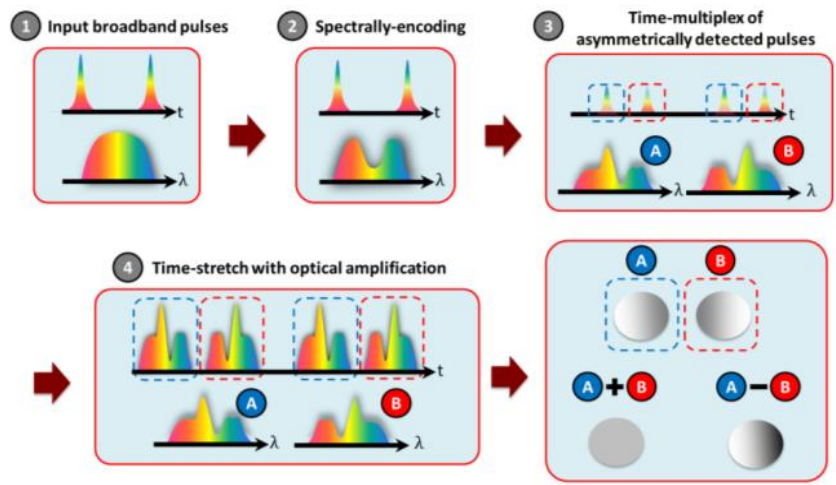

**Figure 2 | General schematic of an ATOM system.** (**a**) A broadband light pulsed beam (The temporal pulse train and the corresponding spectrum are shown in box 1 of **b**) is first spatially dispersed by a diffraction grating to generate a 1D spectral shower. It is then focused by an objective lens for illumination such that different wavelengths are focused on different positions on the sample, i.e. the flowing cells in this example. The spectrally-encoded light double-passes the cell through another objective lens and a mirror. The double-passed spectrally-encoded pulses, which are now encoded with the spatial information of the sample, are then restored back to an undispersed pulsed beam by the grating (see box 2 in **b**). They are further split into two different paths (beams A and B) such that the two beams are coupled into the fiber core with the equal angles but opposite orientations. Such off-axis coupling angles of both beams are controlled by the steering mirrors shown in **a**. One of the beams is time-delayed so that they are multiplexed in time without temporal overlap. The two time-multiplexed signals reveal opposite phase-gradient contrasts because of the opposite off-axis coupling orientations of beams A and B (see box 3 in **b**). The time-multiplexed signals then undergo the time-stretch process and the in-line optical image amplification to perform wavelength-to-time mapping and to compensate the intrinsic loss in the fiber (see box 4 in **b**). A photodetector and a real-time oscilloscope are used for detecting the signals. Note that the time-stretch signal in each pulse period corresponds to one single-shot line-scan. In each line-scan, two different signals, which have the opposite phase-gradient contrasts (A and B), are obtained simultaneously. By calculating the difference of the two signals (A − B) for each line scan, one can obtain a 2D image with *differential* (enhanced) phase-gradient contrast. By calculating the sum of the two signals (A + B), one can on the other hand obtain a 2D image with absorption contrast. A homogenous sphere is depicted in **b** as an example.

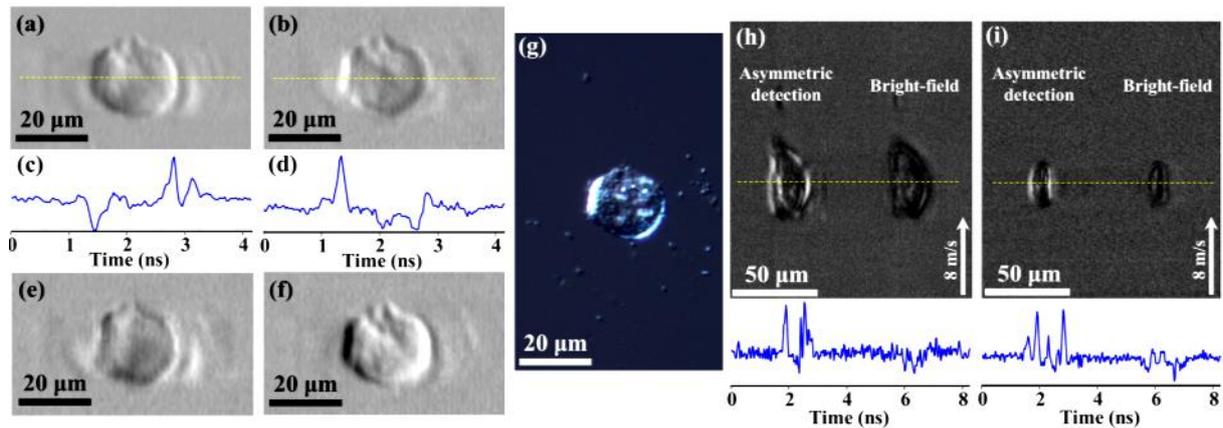

**Figure 3 | Basic performance of ATOM.** (**a,b**) Two ATOM images of a normal hepatocyte cell (MIHA) fixed on a glass slide, which show the opposite phase-gradient contrasts, respectively. (**c,d**) The corresponding line profiles (yellow dotted lines) of the ATOM images in **a,b**, respectively. Each line scan of the image is captured within ~4 ns. (**e**) By calculating the sum of the two ATOM images (i.e. **a** + **b**), an image with absorption contrast can be revealed. (**f**) Differential (enhanced) phase-gradient contrast image can be obtained by calculating the difference of the two images (i.e. **a** - **b**). (**g**) White-light DIC image of the same MIHA cell for comparison. (**h,i**) Demonstration of the contrast enhancement in ATOM by comparing the asymmetric detection scheme (i.e. off-axis fiber coupling) and the bright-field (BF) scheme (i.e. on-axis fiber coupling), in the context of high-speed flow imaging (8 m/s) of the MIHA cells in a PDMS microfluidic channel. 2D images are here acquired by continuous 1D line-scan at a rate of 26 MHz (governed by the repetition rate of the laser), which is naturally provided by the unidirectional cell flow, without any laser beam or sample stage scanning. Two representative time-multiplexed ATOM images are shown in **h** and **i**. In these time-multiplexed ATOM images, they show the image replica of the same cell, as if they are flowing in parallel: one is captured by asymmetric detection (left) whereas another is captured by on-axis detection, i.e. BF time-stretch images (right). The bottom insets are the corresponding line profiles (yellow dotted lines) of ATOM images and BF time-stretch images shown in **h,i**. Note that in one single-shot line-scan, both phase-gradient and BF contrasts of the same cell are captured within ~8 ns.

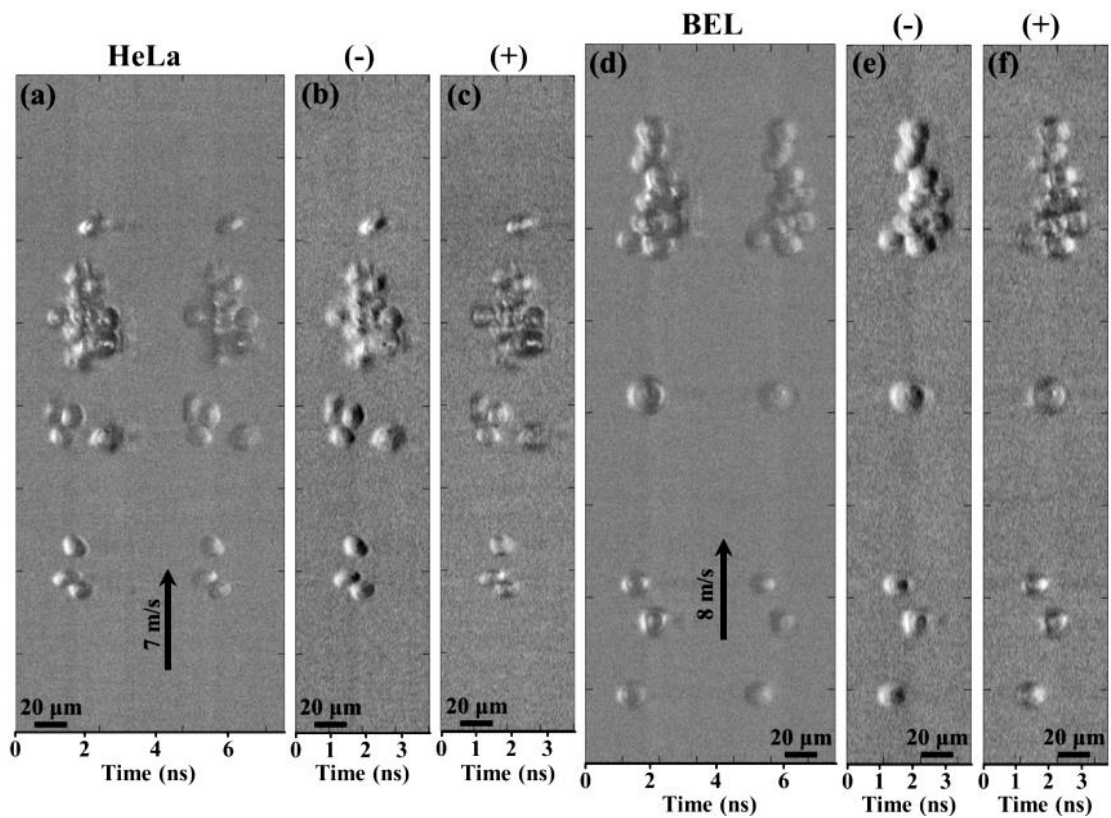

**Figure 4. Imaging of mammalian cells by ATOM in ultrafast flow.** (**a**) Time-multiplexed ATOM images of stain-free HeLa cells flowing at a speed of 7 m/s. In this time-multiplexed image, it appears that two groups of identical cells are flowing in parallel. The left group represents the image replica, but with opposite phase-gradient contrast, of the right group. It can be seen that the shadow casts in these two groups of cells aside on the opposite side of the individual cells. (**b**) Differential (enhanced) phase-gradient-contrast ATOM image of the same cell group. It is done by calculating the difference of the two images (i.e. left and right groups) shown in **a**. (**c**) Absorption-contrast ATOM image of the same cell group, which is done by calculating the sum of the two images (i.e. left and right groups) shown in **a**. More examples of the HeLa images are shown in Supplementary Fig. S6. (**d**) Time-multiplexed ATOM images of stain-free BEL-7402 cells flowing at a speed of 8 m/s. (**e**) Differential (enhanced) phase-gradient-contrast and (**f**) absorption-contrast ATOM images the same cell group, respectively. More examples of the BEL-7402 images are shown in Supplementary Fig. S7. Note that the strength of ATOM is particularly exemplified in these images: the individual cells in the aggregates are clearly distinguishable, that would be otherwise regarded as a single entity in a conventional non-imaging flow cytometer and thus produces a higher rate of false positive in cell screening/identification.

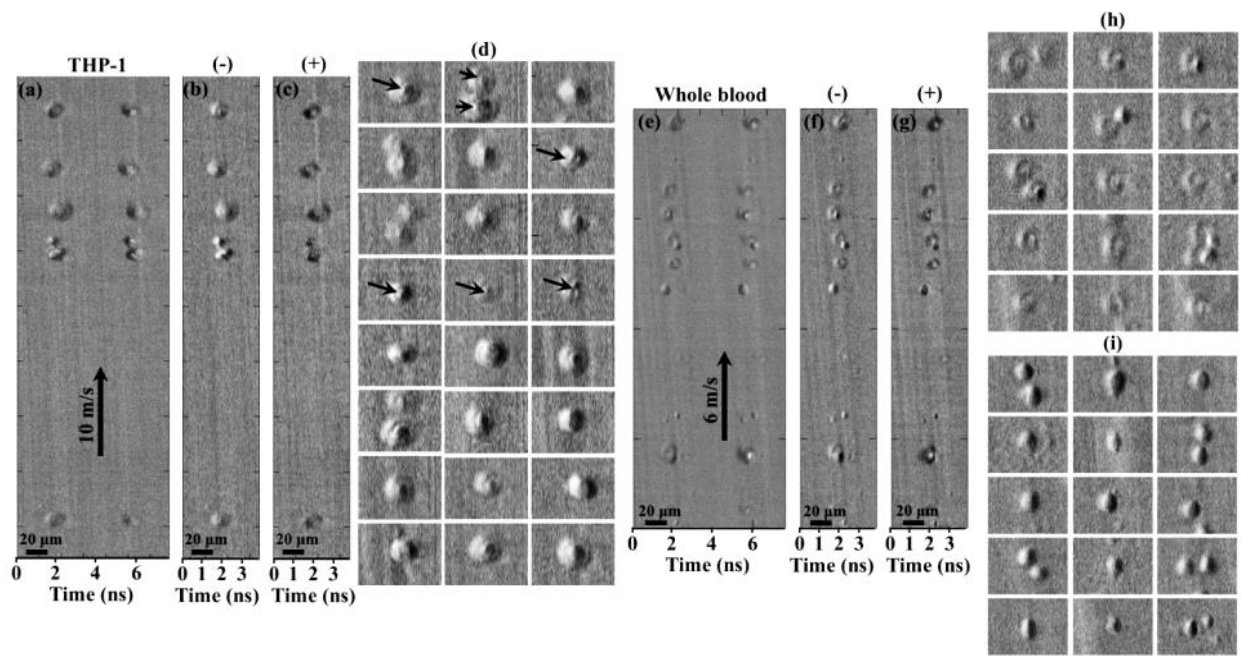

**Figure 5 | Imaging of acute monocytic leukemia cells (THP-1) and normal human blood cells (from whole blood) by ATOM in ultrafast flow.** (**a**) Time-multiplexed ATOM images of the stain-free THP-1 cells flowing at a speed of 10 m/s. (**b**) Differential (enhanced) phase-gradient-contrast ATOM image obtained by subtraction of the two opposite-contrast images (the left and right group of cells) in **a**. (**c**) Absorption-contrast ATOM image obtained by addition of the two opposite-contrast images (the left and right group of cells) in **a**. More differential phase-gradient-contrast ATOM images of THP-1 cells are shown in **d**. More examples are shown in Supplementary Fig. S8. The enhanced contrast enables visualization of the nuclei of the cells (indicated by the arrows). (**e**) Time-multiplexed ATOM images of stain-free blood cells (freshly from whole blood) flowing at a speed of 6 m/s. (**f**) Differential (enhanced) phase-gradient-contrast ATOM image. (**g**) Absorption-contrast ATOM image. More examples are shown in Supplementary Fig. S9. The enhanced phase-gradient-contrast ATOM images of the red blood cells (RBCs) clearly reveal their classical biconcave disk shape, as shown in **h**. It can also help visualizing the swelled RBCs in the whole blood cells, which show spherical or elliptical shapes (consistent with the inspection under the ordinary DIC microscope, see Supplementary Fig. S9).

# Supplementary Information of "Asymmetric-detection time-stretch optical microscopy (ATOM) for ultrafast high-contrast cellular imaging in flow"

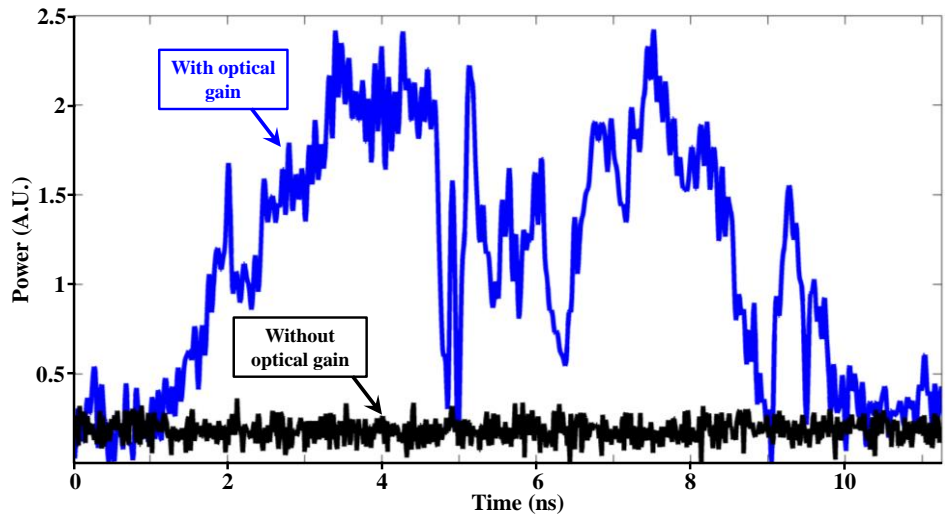

**Supplementary Figure 1 | The significance of the in-line optical amplification in ATOM:** (Black) Without in-line optical amplification, the time-stretch signal is invisible and is buried below the noise floor of the detection system (i.e. photodetector) mainly due to the intrinsic loss of GVD in the dispersive fiber. (Blue) With the optical amplification (20 dB on-off gain in this case), the amplified time-stretch signal is brought above the noise floor and becomes detectable. Viable options for the in-line optical amplification can be discrete amplifiers (namely SOA, ytterbium-doped fiber amplifier, or optical parametric amplifier), or distributed amplifiers, such as distributed fiber Raman amplifier.

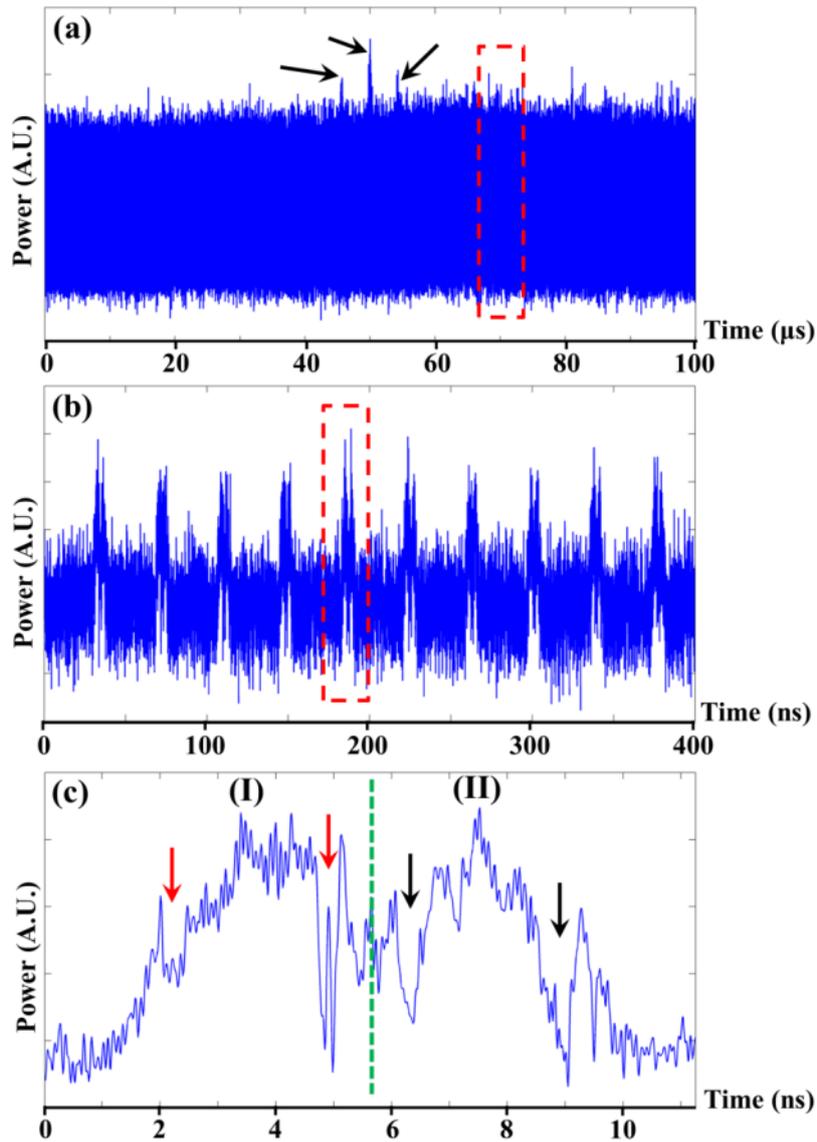

**Supplementary Figure 2 | Raw time-stretch data stream captured by ATOM.** (**a**) Raw time-stretch pulse train obtained by the real-time oscilloscope. Arrows indicate the captured cell events. (**b**) Zoom-in view of the pulse trains in dashed box in **a**. Each period represents a single line scan of an ATOM image which consists of two time-multiplexed scans revealing the opposite phase-gradient contrast (because of the opposite off-axis fiber coupling angles). (**c**) Zoom-in view single time-stretch waveform in dashed box in **b**. The pulse consists of two time-multiplexed ATOM line scans captured with the opposite fiber-coupling angles. Green dashed line separates the two time-multiplexed pulses (labeled as I and II). Red and black arrows indicate channel wall positions in each time-stretch pulse I and II, respectively.

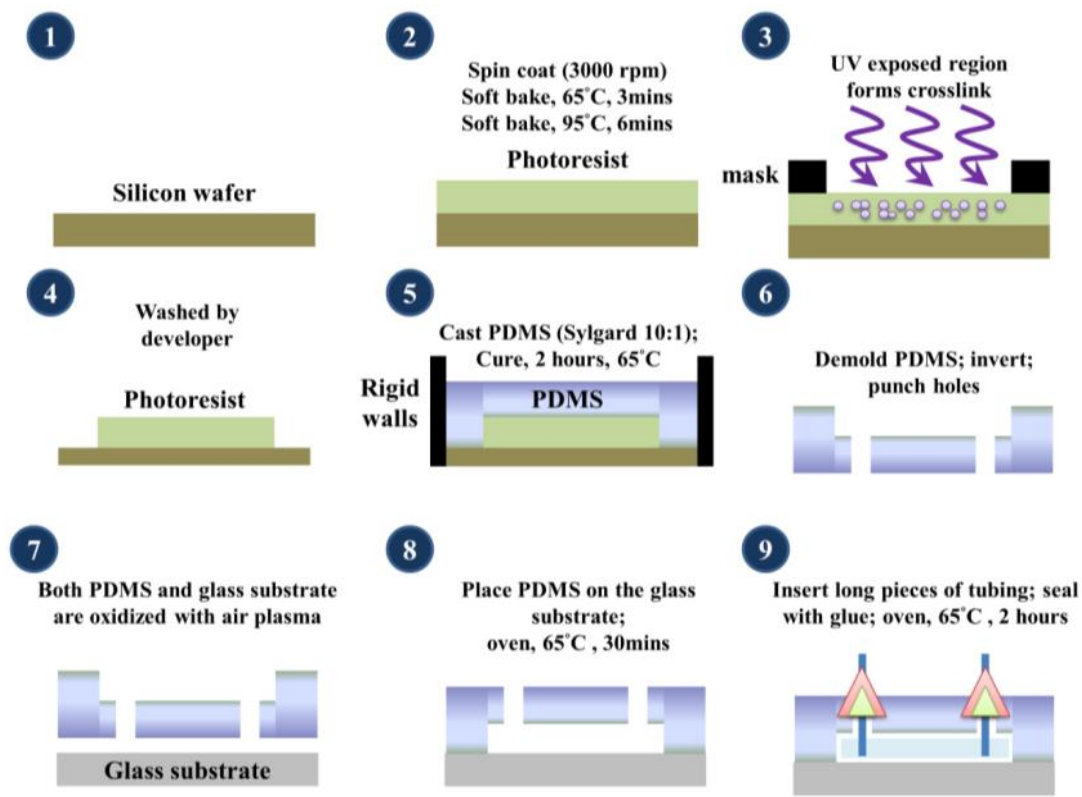

**Supplementary Figure 3 | Fabrication steps of the microfluidic channel.** The microchannel is fabricated by curing Polydimethylsiloxane (PDMS) on a silicon mold. The silicon mold is prepared by soft lithography. First, a layer of photoresist (MICRO CHEM SU-8 2025) is coated on a silicon wafer using a spin coater (Apex Instruments Co. spinNXG-P1). Then, the photoresist is soft baked at 65 and 95 $^{o}$C for 3 and 6 minutes on a hot plate, respectively. The silicon wafer and photoresist are then cooled under ambient temperature before being exposed to ultraviolet (UV) light. A maskless soft lithography machine (Intelligent Micro Patterning, LLC SF-100 XCEL) is used to transfer a CAD pattern onto the photoresist. Four seconds of exposure time is chosen. The exposed photoresist is post-baked for 1 and 6 minutes at 65 and 95$^{o}$C on the hot plate, respectively. Afterwards, the photoresist undergoes a developing step by immersing the photoresist-coated silicon wafer in SU-8 developer (MICRO CHEM) for 5 minutes. The silicon wafer is rinsed with isopropyl alcohol followed by a drying step of blow either nitrogen or air over the wafer. The PDMS precursor (SYLGARD 184 Silicone Elastomer) is mixed with the corresponding curing agent with a ratio of 10:1 before pouring onto the silicon mold. A Polymethylmethacrylate block is designed and placed on the silicon wafer to achieve thin PDMS in the imaging section of the channel (see Supplementary Fig. 4(a)). The PDMS is cured in an oven at 65$^{o}$C for 2 hours. Then, the PDMS is demolded and holes are punched using a biopsy punch (Miltex 33-31AA). The PDMS and a glass slide are washed before they are exposed to oxygen plasma (Harrick Plasma PDC-002). This process makes the surface hydrophilic and enhances the bonding strength between the glass slide and the PDMS block. Afterwards, the whole device is placed in an oven at 65$^{o}$C for 30 minutes for better bonding performance. Lastly, sections of plastic tubing (Scientific Commodities, Inc. BB31695-PE/2) is inserted into the punched holes with PDMS glue applied at the gap between the tubing and the PDMS block to prevent water leakage.

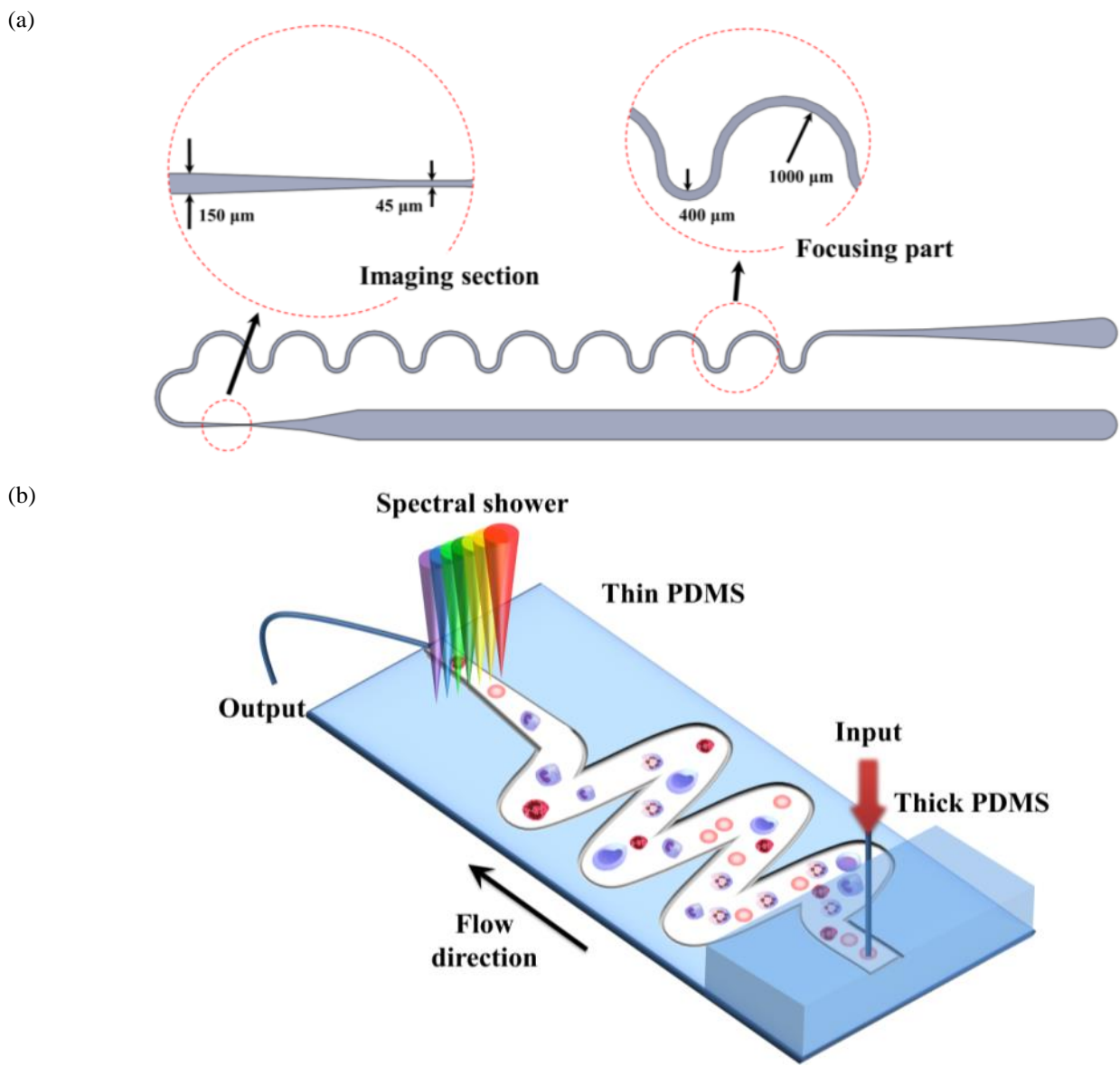

**Supplementary Figure 4 | Schematic of the microfluidic channel platform for ATOM.** (**a**) Detailed channel structure drawn to scale. The channel is divided into a focusing part and an imaging section. The focusing part of the device helps focus the objects in the channel along the same streamline to facilitate subsequent imaging. The positions of the individual objects in the channel can be tuned by adjusting the balance between the inertial lift force and the viscous drag force through changing the fluid flow rate. (**b**) The focused objects are fed into the imaging section of the device, which consists of a microfluidic channel in a thinner PDMS slab (thick PDMS: ~4 mm, thin PDMS: ~1.5 mm). The spectral shower is illuminated in the imaging section, with the spectral encoding direction perpendicular to the microfluidic flow direction. The flow speed in the microfluidic channel is as high as ~10 m/s, which is limited only by the pressure that the channel can withstand. In the imaging of cells, this flow speed corresponds to an imaging throughput up to ~100,000 cells/sec,

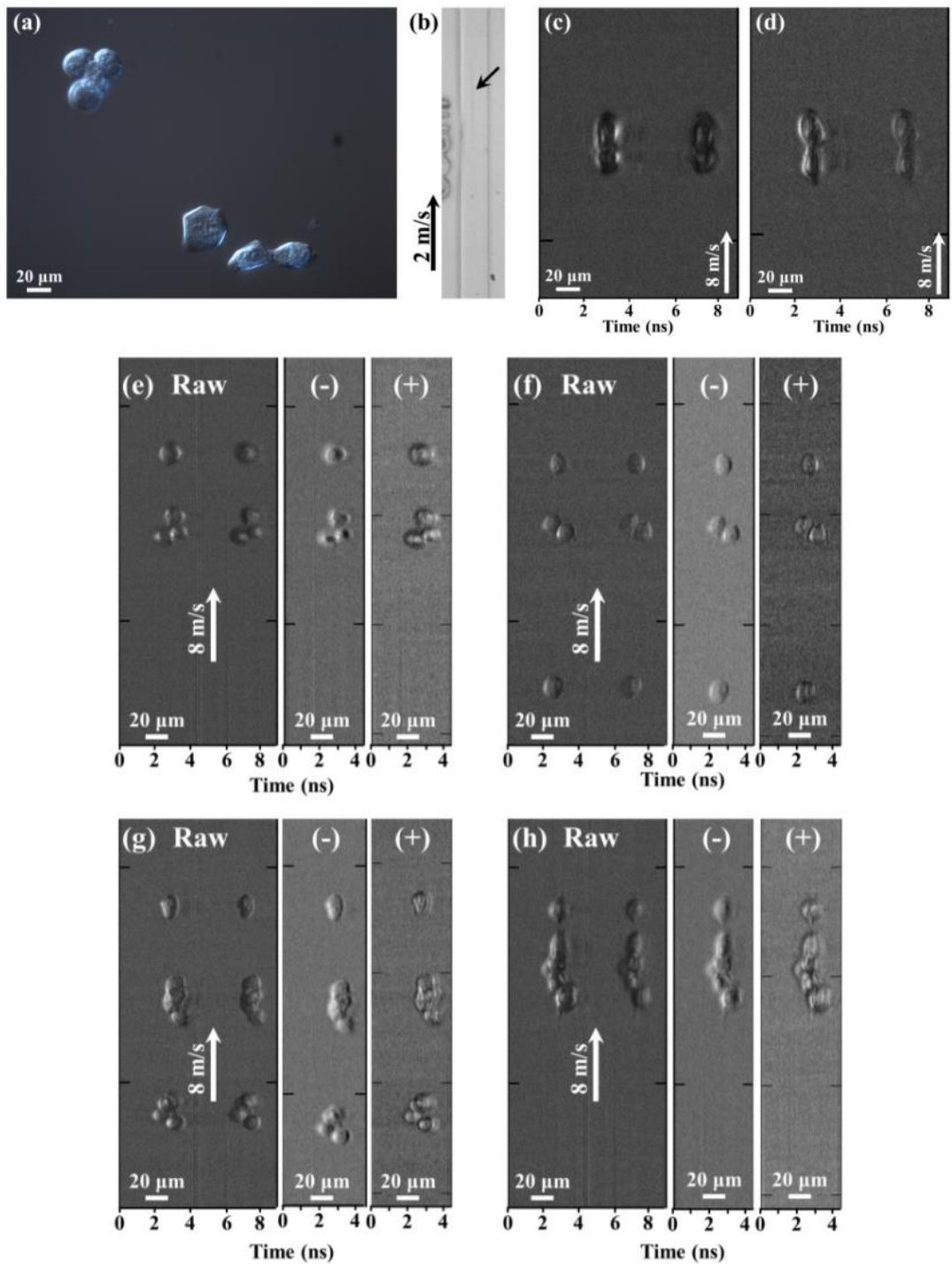

**Supplementary Figure 5 | Imaging of the MIHA cells flowing in the microfluidic channel.** (**a**) White-light DIC image of stain-free MIHA cells on glass slide. (**b**) Image of the MIHA cells flowing at a speed of 2 m/s, captured by a high speed CMOS camera (frame rate of 19000 fps). Black arrow indicates MIHA cell which appears to be completely blurred in the CMOS image. This is in great contrast to ATOM images, which are all motion-blur-free and are captured with the ultrafast flow speed of 8 m/s. (**c,d**) High-speed flow (8 m/s) imaging of the budding MIHA cells by ATOM. (**c,d**) In these time-multiplexed ATOM images, they show the image replica of the same cell, as if they are flowing in parallel: one is captured by asymmetric detection (left) whereas another is captured by on-axis detection, i.e. BF time-stretch images (right). (**e-h**) High-speed flow MIHA images by ATOM. **Left:** Time-multiplexed ATOM images of stain-free MIHA cells flowing at 8 m/s. **Middle:** Differential phase-gradient-contrast ATOM images (–). **Right:** Absorption-contrast ATOM images (+).

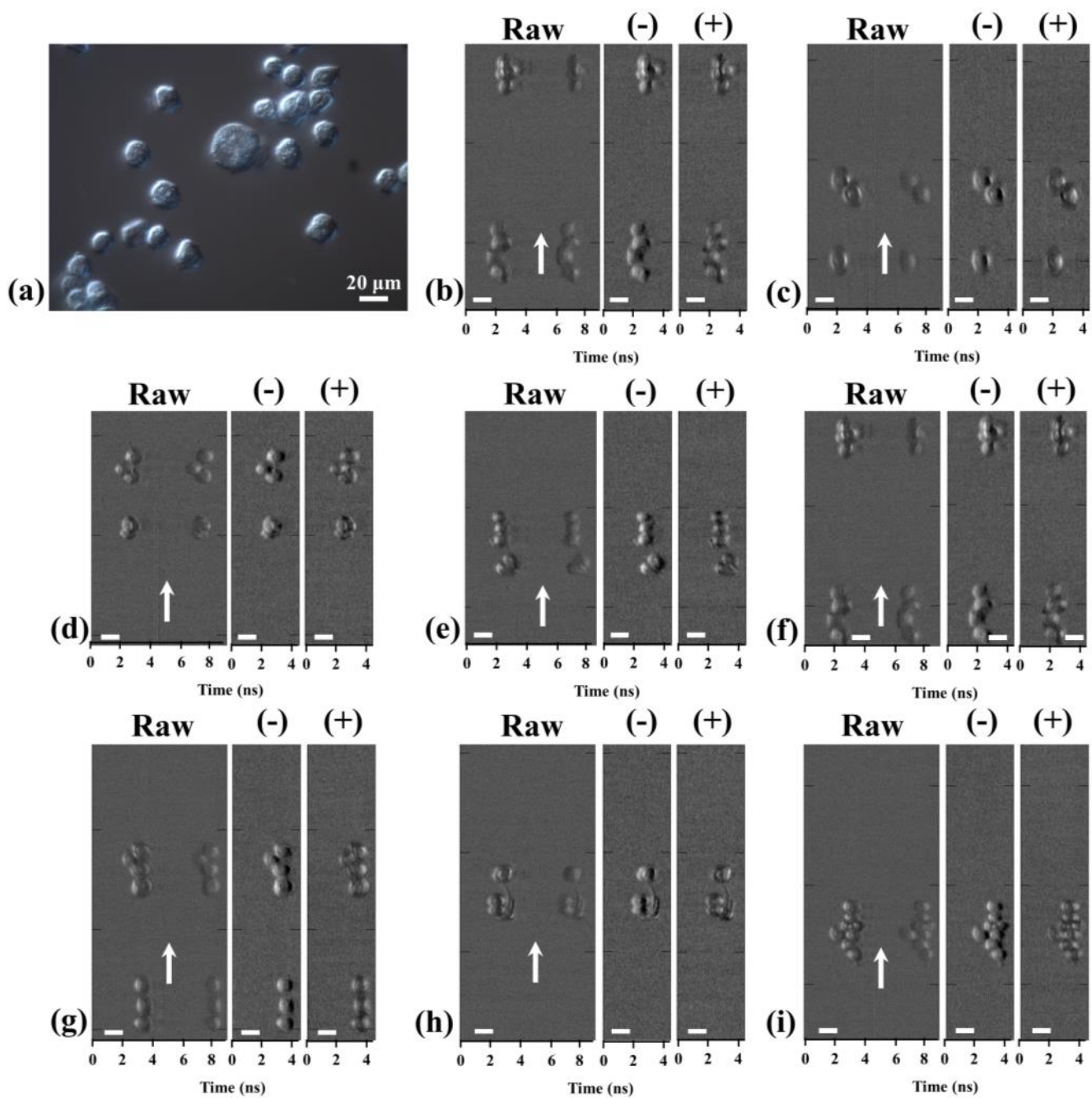

**Supplementary Figure 6 | Imaging of BEL cells flowing in the microfluidic channel.** (**a**) White-light DIC image of stain-free BEL cells on glass slide. (**b-i**) ATOM images of BEL cells flowing in the microfluidic channel at a speed of 8 m/s. The arrows indicate the flow direction. **Left:** Time-multiplexed ATOM images of stain-free BEL cells flowing at 8 m/s. **Left:** Time-multiplexed ATOM images of stain-free BEL cells flowing at 8 m/s. **Middle:** Differential phase-gradient-contrast ATOM images (–). **Right:** Absorption-contrast ATOM images (+). The scale bars represent 20 μm.

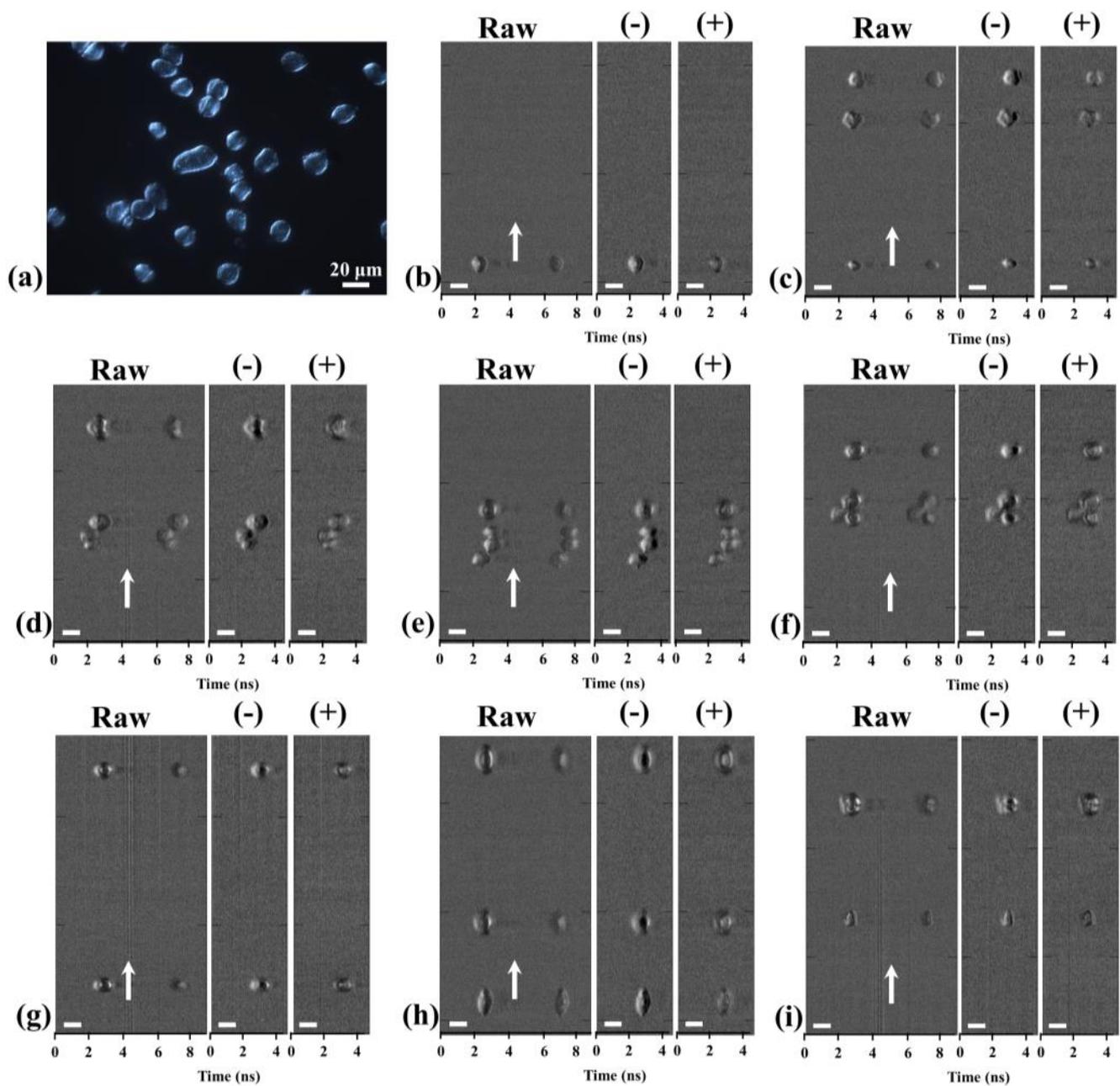

**Supplementary Figure 7 | Imaging of HeLa cells flowing in the microfluidic channel.** (**a**) White-light DIC image of stain-free HeLa cells on glass slide. (**b-i**) ATOM images of HeLa cells flowing in the microfluidic channel at a speed of 8 m/s. The arrows indicate the flow direction. **Left:** Time-multiplexed ATOM images of stain-free HeLa cells flowing at 7 m/s. **Left:** Time-multiplexed ATOM images of stain-free HeLa cells flowing at 7 m/s. **Middle:** Differential phase-gradient-contrast ATOM images (–). **Right:** Absorption-contrast ATOM images (+). The scale bars represent 20 μm.

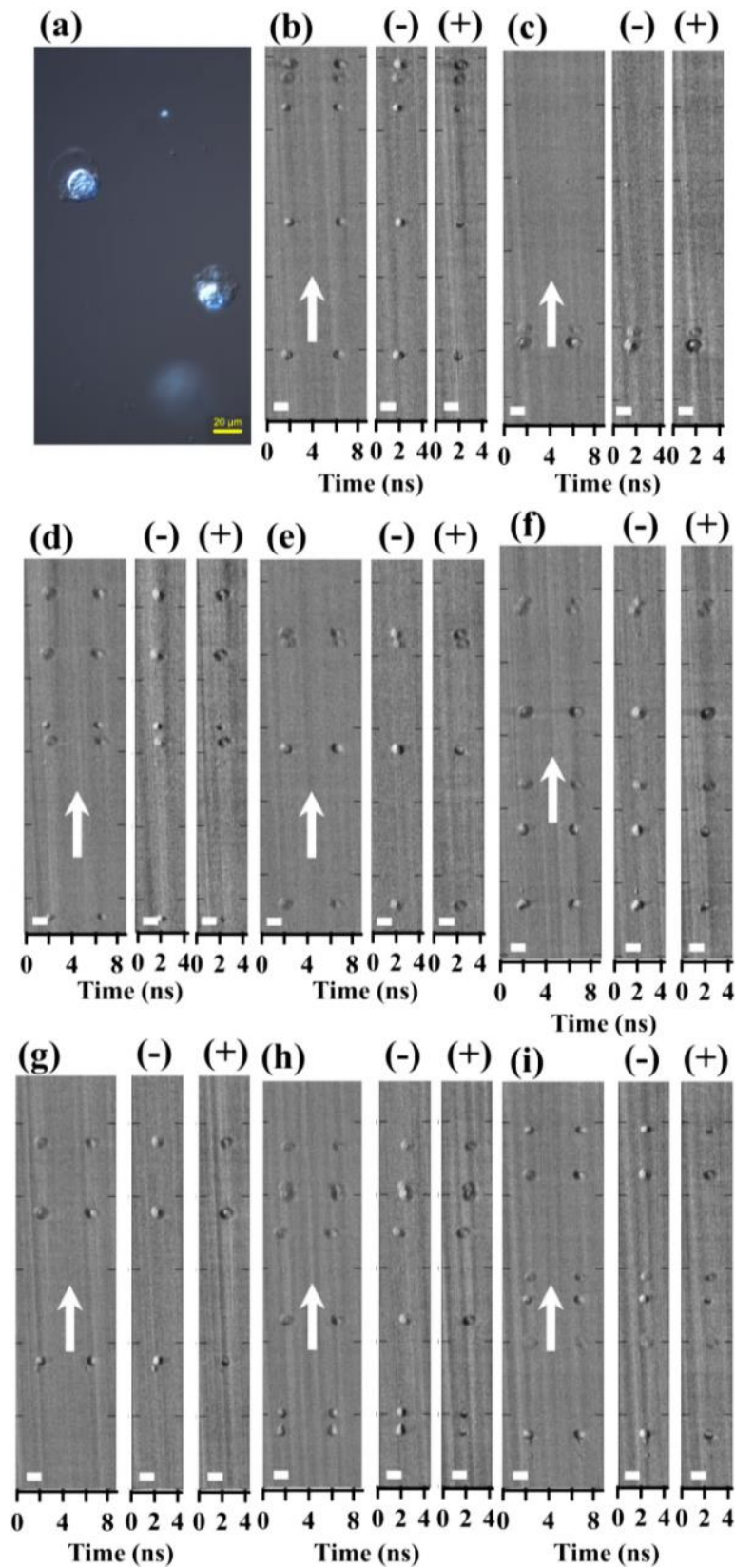

**Supplementary Figure 8 | Imaging of acute monocytic leukemia cells (THP-1).** (**a**) White-light DIC image of stain-free THP-1 cells on glass slide. (**b-i**) ATOM images of THP-1 cells flowing in the microfluidic channel at a speed of 10 m/s. The arrows indicate the flow direction. **Left:** Time-multiplexed ATOM images of stain-free THP-1 cells flowing at 10 m/s. **Left:** Time-multiplexed ATOM images of stain-free THP-1 cells flowing at 10 m/s. **Middle:** Differential phase-gradient-contrast ATOM images (−). **Right:** Absorption-contrast ATOM images (+). The scale bars represent 20 μm.

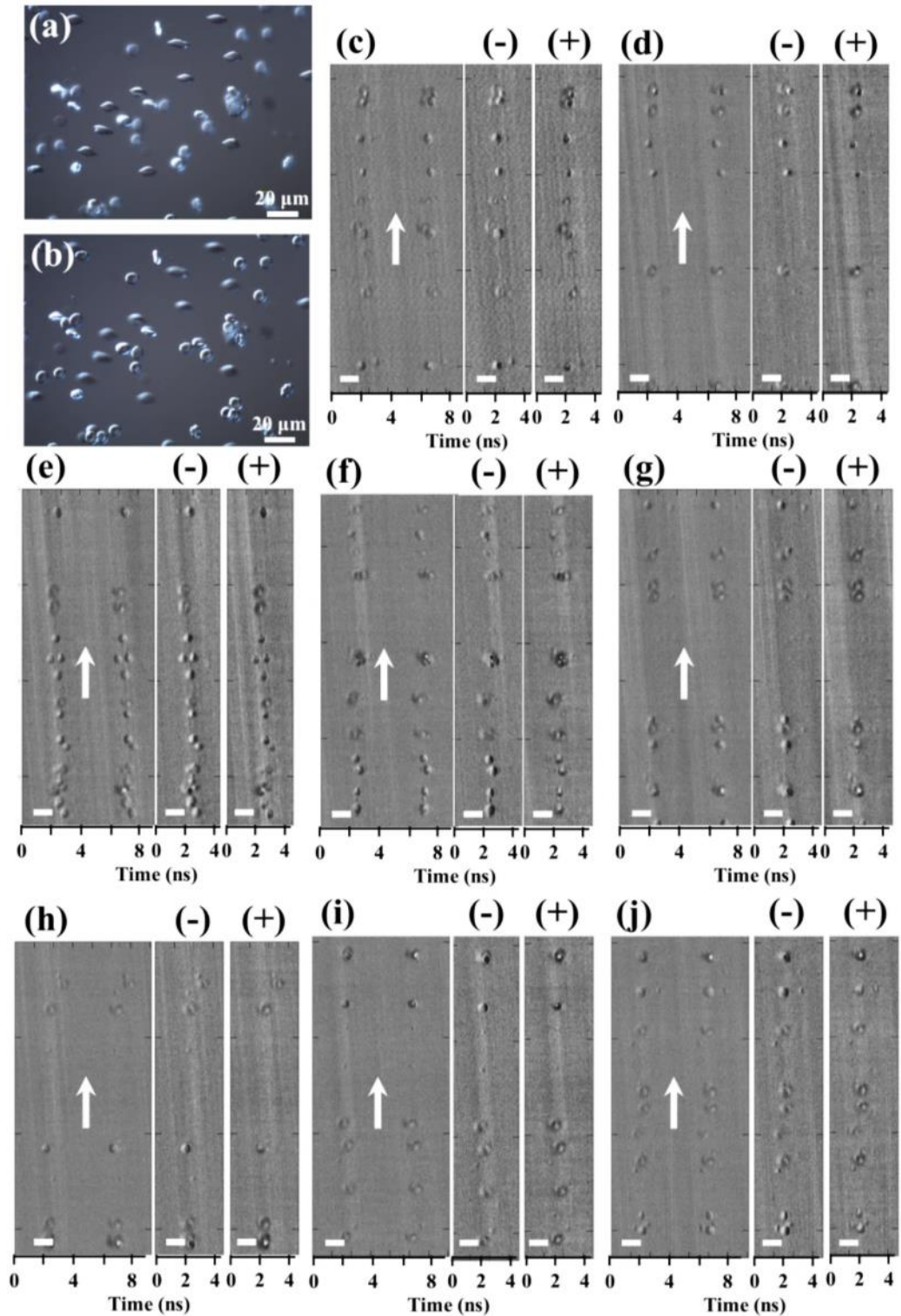

**Supplementary Figure 9 | Imaging blood cells from whole blood.** (**a-b**) White-light DIC image of stain-free blood cells (obtained from whole blood) on glass slide. (**a**) and (**b**) shows the blood cells at two different focusing planes in order to visualize both the biconcave-disk-shaped red blood cells (RBCs) as well as the swelled RBCs. (**b-j**) ATOM images of blood cells flowing in the microfluidic channel at a speed of 6 m/s. The arrows indicate the flow direction. **Left:** Time-multiplexed ATOM images of stain-free blood cells flowing at 6 m/s. **Left:** Time-multiplexed ATOM images of stain-free blood cells flowing at 6 m/s. **Middle:** Differential phase-gradient-contrast ATOM images (–). **Right:** Absorption-contrast ATOM images (+). The scale bars represent 20 μm.